\documentclass{elsart}
\usepackage{amssymb}
\usepackage{amsmath}
\usepackage{epsfig}
\begin{document}
\begin{frontmatter}

\title{Study of the Urban Road Networks of Le Mans}

\author[a,b]{J. Jiang \corauthref{email}}
\corauth[email]{Corresponding author. \\
\textit{E-mail address}: jjiang@ismans.fr\\}
\author[a]{, M. Calvao}
\author[a]{, A. Magalhases}
\author[a]{, D. Vittaz}
\author[a]{, R. Mirouse}
\author[a]{, F. Kouramavel}
\author[a]{, F. Tsobnange}
\author[a]{, and Q.A. Wang}

\address[a]{Institut Sup\'erieur des Mat\'eriaux du Mans,
44, Avenue F.A. Bartholdi, 72000 Le Mans, France}
\address[b]{Complexity Science Center, Institute of
Particle Physics, Hua-Zhong (Central China) Normal University,
Wuhan 430079, China}

\begin{abstract}
An urban road network of Le Mans in France is analyzed. Some
topological properties of network are investigated, such as degree
distribution, clustering coefficient, diameter, and characteristic
path length. These results suggest that our network is a "small-
world" network with short average shortest path and large
clustering coefficient. Furthermore, double power-law distribution
is found in degree distribution which is distinct from the single
power-law and a novel degree distribution function is also given.
Some analysis on this function extend the comprehension of the
origination of the double power-law distribution widely dispersed
in nature.

\vfill
\end{abstract}

 \maketitle
\end{frontmatter}

\def\tc{T_{\rm cr}}
\section{Introduction}
During the last decade, the growing availability of large
detabases, the increasing computing powers, as well as the
development of reliable data analysis tools, have constituted a
better machinery to explore the topological properties of complex
networks \cite{bara,newman,rpas,watts}. Most of studies has
revealed that, despite the inherent differences, main real
networks are characterized by the same topological properties,
such as relatively small path lengths, high clustering
coefficients, scale-free degree distributions, degree
correlations, motifs, and community structures. As a particular
class of complex networks, spatial networks are different from
other networks, which are those embedded in the real space, i.e.
networks whose nodes occupy a precise position in two or three
dimensional Euclidean space, and whose edges are real physical
connections. It is not surprising that the topology of spatial
networks is strongly constrained by their geographical embedding,
such as airport networks, urban street networks. In this paper we
focus on an urban road networks.

Urban road networks, in one way or another, are portrayals of the
history of the country's development. Hence the topology studies
of various spatial networks could show significantly different
degree distributions. For example, the degree distribution of
internet network (presupposing that the network nodes are routers)
has the form $P(k)\sim k^{-\gamma}$ where
$\gamma_{out}=\gamma_{in}\approx2.48$ \cite{bara}. The power grid
of the western US is described by a complex network with
exponential form degree distribution \cite{amaral}. In addition,
the analysis of the spatial distribution of the network nodes
shows that both indicated networks are fractals with the method of
box counting \cite{rpas, gast}. But, here, we show other distinct
degree distribution, double power-law, in urban road networks of
Le Mans in France (Le Mans is one small city in France). Urban
road networks with links and nodes representing road segments and
junctions respectively which is a primal graph, exhibit unique
features different from other networks. As road networks are
almost planar, they show a very limited range of node degrees and
could never be scale-free like airport networks or the internet
\cite{gast}. Nevertheless, there is an interesting connection
between those scale-free spatial networks and road networks, since
both are extreme cases of an optimization process minimizing
average travel costs along all shortest paths, given a set of
nodes and a total link length.

In the present paper, the organization is as follows. Section 2
presents the results of degree distribution and cumulative degree
distribution of urban road network of Le Mans. In section 3, we
analyze the relationship between the double power-law distribution
function and its parameters. In section 4 and 5, some topological
properties of network are investigated, such as clustering
coefficients, diameter and average shortest path length.
Conclusions and discussions are given in the last part, section 6.

\section{Degree distribution}

We investigate the urban road network of Le Mans in France.
Previous works on urban traffic network usually were based on a
primal graph, where intersections were turned into nodes and roads
into edges. Here, we look at them with other paradigm, a dual
graph, where a city is transformed into a topological graph by
mapping the roads into nodes and the intersections into edges
between the nodes. The advantage of the dual graph is that it does
not exhibit any geographic constrain compared to primal graph.
Hence, in structure of our network, there are 1585 nodes and 5066
edges.

\begin{figure} [h]
\includegraphics[scale=.65]{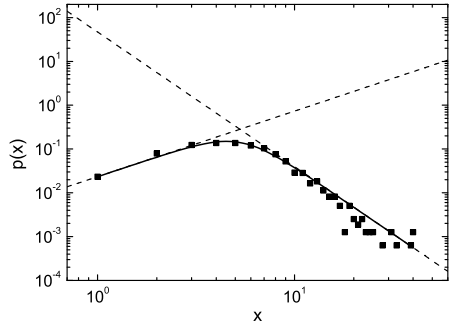}
\centering \caption{Degree distribution of urban road network of
Le Mans city in France. $p(x)$ is the degree distribution
function, and $x$ is the degree of each node. ``double power-law"
phenomenon is obviously showed in log-log plot and our fitting
curve with black solid line matches the data well. Two straight
dashed lines define two power laws with scale exponents 3.1 and
-1.5 respectively. Our fitted function is
$p(x)=46.7/(x^{3.1}+1999*x^{-1.5})$. } \label{pxfig}
\end{figure}

During the research of complex network science, degree
distribution ($p(x)$, $x$ is the degree of one node) is always one
of the most important characteristics. In Figure \ref{pxfig}, we
plot the degree distribution in double-log plot and fit the data
with black solid line with function as follows, which is called
``double power-law" function by us:
\begin{equation}     \label{p}
p(x)=a/(x^b+c*x^d),
\end{equation}
where $a=46.7,b=3.1,c=1999,d=-1.5$. The slopes of two straight
dashed lines are 3.1 and -1.5 respectively. The reason for
choosing this kind of function was explained in \cite{liwei}. Many
similar distributions were found in other researches including the
highway networks of Korean \cite{Jung} and urban public transport
networks of Los Angeles in American \cite{Ferber}. Nevertheless,
they were named ``truncated power-law distributions" by authors.
In fact, abundant works with ``double power-law" were analyzed as
the truncated power law \cite{Kalapala, Namikawa, Niwa, Biham,
Mossa, Gupta, Gupta2} in which more attention was paid to large
degree value in one power-law region and small degree value in
another one was neglected with various reasons. In addition, it is
worth to notice that in Figure \ref{pxfig}, the node with degree 5
has the largest probability, in other words, crossover point has
maximum value of probability, which is similar to the Moscow
region road network in which the node with degree 3 had largest
probability in degree distribution \cite{moscow}. It is very
different from non-road spatial network's degree distributions in
which the node with smallest degree, not crossover point, has the
largest probability generally. This is decided by the feature of
spatial networks such as transportation networks, since the
topology of spatial networks is strongly constrained by their
geographical embedding. Furthermore, both belonging to
transportation networks, road network still has distinct degree
distribution form from airport networks \cite{liwei} in which
crossover point does not have maximum value of probability. Since
there is a strong restriction to the growth of the degree in road
networks which is induced by the geographical arrangement of the
edges, however, there is no so strong geographical restriction to
the edges existence in the airport networks.

Towards degree distribution function, combined Eq.~(\ref{p}) and
Figure \ref{pxfig}, we could simply find that parameter $a$ is the
normalization constant, $b$ and $d$ are two scale exponents in two
regions of power-law respectively. So, in principle, from the
standpoint of mathematics, this kind of function could produce any
form of ``double power-law" distribution, which could be seen
having some relationships with the origin of this phenomenon.

\begin{figure} [h]
\includegraphics[scale=.65]{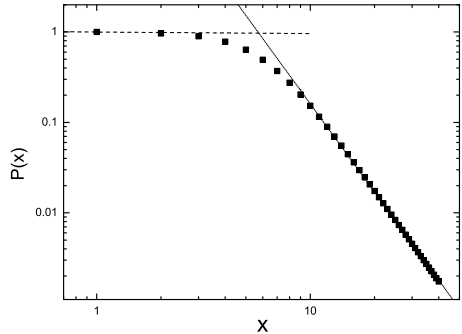}
\centering \caption{Cumulative distribution function of degree of
urban road network of Le Mans city in France. $P(x)$ is the
cumulative degree distribution function, and $x$ is the degree of
each node. ``double power-law" phenomenon is also showed in
log-log plot. Two straight dashed lines define two power laws with
scale exponents -3.35 and -0.017 respectively.} \label{cumupxfig}
\end{figure}

An alternative way of presenting degree data is to make a plot of the
cumulative distribution function
\begin{equation}     \label{P}
P(x)=\int_x^{40} a/(x^b+c*x^d)dx,
\end{equation}
which is the probability that the degree is greater than or equal to
$x$ and 40 is the maximum degree in our network. The advantage of
cumulative distribution is that all the original data are represented
and it reduces the noise in the tail. After calculation of Eq.~(\ref{P}),
we get
\begin{equation}
P(x)=\frac{-42.5+0.4*x^{2.5}Hpergeometric2F1[0.54,1,1.54,-0.0005*x^{4.6}]}{x-40},
\end{equation}
where $Hypergeometric2F1[a,b,c,z]$ is the hypergeometric function defined by:
\begin{eqnarray}
{}_2F_1(a,b;c;z)=\frac{{\Gamma(c)}}{{\Gamma(b)\Gamma(c-b)}}\int_0^1
{t^{b-1}}(1-t)^{c-b-1}(1-tz)^{-a} dt.
\end{eqnarray}
From Figure \ref{cumupxfig}, we also find that there is double
power-law distribution in log-log plot with two exponents -0.017
and -3.35 respectively.

\section{Relationship between double power-law function and its parameters}
We make an elaborate analysis on the expression of $p(x)$.
Firstly, we do the normalization for the probability function by
using equation $ \sum\limits_{x = 1}^{100} {p(x) = 1}$ since the
degree of each node is discrete variable. As parameter $a$ is the
normalization constant, so we pay more attention to the influence
on the expression of $p(x)$ caused by the varying of parameters
$b,c,d$. From the Figure \ref{pxfig}, we could find that $b$ and
$d$ have the directly relationship with the exponents of two
power-law regions. Hence, our discussions mainly focus on the
varying of $b$ and $d$. Based on the different features of degree
distribution in airport networks and urban road networks, our
discussions are divided into two classes: one is that degree at
crossover point has non-largest probability, behaved as both of
$b,d$ are positive like airport networks; another is that degree
at crossover point has largest probability, behaved as either $b$
or $d$ is negative, the rest is positive, like urban road
networks. The case of $b=d$ reduces to familiar single power-law
distribution. In each class, we attempt to analyze the role of
each parameter played. At the same time, we make sure that $p(x)$
is always normalized. We define the intersection of two power-law
in all figures as ``crossover" and define the scaling behavior in
the region of $1<x<x_{crossover}$ as ``the first power-law" and
define that in the region of $x_{crossover}<x<100$ as ``the second
power-law".

\begin{figure} [h]
\includegraphics[scale=.45]{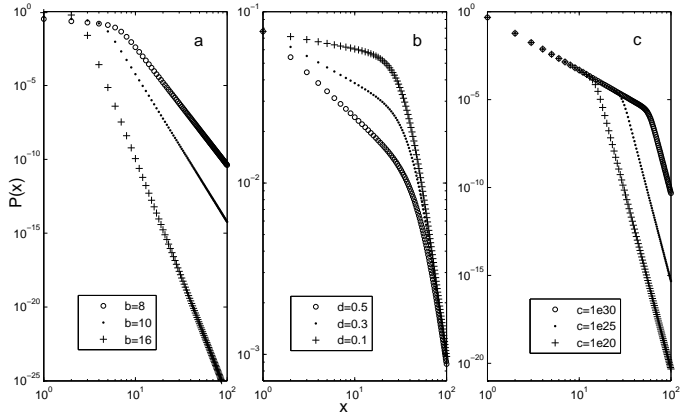}
\centering \caption{Variation tendency of probability function
$p(x)$ against its parameters $b, c, d$ when $b>d>0$ in log-log
plot. In the panel $a$, the exponents of the first power-law are
identical since parameter $d$ is fixed. Those of the second
power-law become bigger with $b$ increasing. While the value of
crossover becomes smaller. In the panel $b$, the exponents of the
second power-law are identical because of invariance of parameter
$b$, those of the first power-law behave the same as panel $a$.
But the value of crossover increases along with parameter $d$. In
the panel $c$, the exponents of each power-law are identical
respectively by reason of unchanged $b$ and $d$. Parameter $c$
directly influences the position of the crossover.}
\label{pbggdfig}
\end{figure}

In the first case $b>d>0$, we choose the probability function in
the form of $p(x)=a/(x^b+1293139*x^{0.5})$ which is a monotonical
function. In the panel $a$ of Figure \ref{pbggdfig}, it is found
that the exponents of the first power-law are the same with fixed
$d$. As $b$ increases, the exponents of the second power-law
become bigger and the value of crossover decreases. In the panel
$b$, the expression of function is $p(x)=a/(x^4+1293139*x^d)$. Due
to the identity of parameter $b$, three curves have the same
exponents of the second power-law. As $d$ increases, the exponents
of the first power-law become bigger and the value of crossover
increases. In the panel $c$, we have the function
$p(x)=a/(x^{20}+c*x^3)$. As the result of equality of both $b$ and
$d$, the overlap of curves of the first power-law and the
parallelity of curves of the second power-law are showed
evidently. The effect of parameter $c$ is to change the position
of crossover. The value of crossover increases with parameter $c$
increasing. So, based on above results, we draw a conclusion that
when $b>d>0$, parameter $b$ and $d$ are directly in charge of the
exponents of the second and the first power-law respectively.
Parameter $c$ is in charge of the position of crossover when $b$
and $d$ are fixed.

\begin{figure} [h]
\includegraphics[scale=.45]{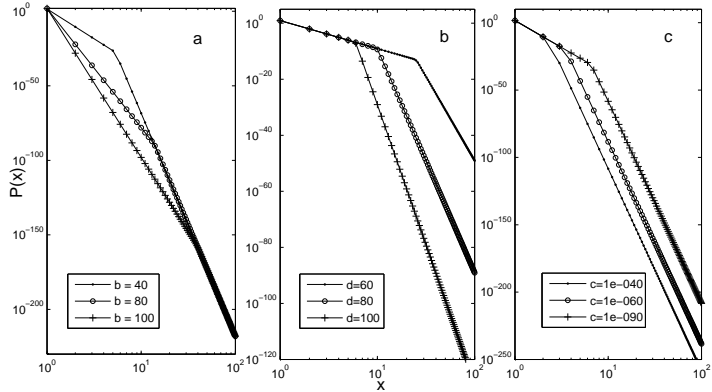}
\centering \caption{Variation tendency of probability function
$p(x)$ against its parameters $b, c, d$ when $0<b<d$ in log-log
plot. In the panel $a$, the exponents of the second power-law are
identical since parameter $d$ is fixed. Those of the first power-
law become bigger along with parameter $b$. In the meantime, the
value of crossover varies bigger. In the panel $b$, the exponents
of the first power-law are identical because of unchanged
parameter $b$. Those of the second power-law become bigger along
with $d$. The value of crossover varies smaller when $b$
increases. In the panel $c$, the exponents of each power-law are
the same respectively. The value of crossover decreases when
parameter $c$ increases, which behaves opposite to that in Figure
\ref{pbggdfig}. } \label{pblldfig}
\end{figure}

In the second case $0< b<d$, we choose
$p(x)=a/(x^b+10^{-80}*x^{150})$. In Figure \ref{pblldfig}, we make
a comparison with the Figure \ref{pbggdfig}. It is observed that
the variation tendency in the panel $a$ and $b$ of Figure
\ref{pblldfig} is quite the reverse of them in Figure
\ref{pbggdfig}. The variation tendency in the panel $c$ of both
figures are mostly the same. So, from discussions of the first
class , it could be seen that in networks such as airport
networks, the probability function is monotonical decreasing. In
addition, the first power-law with smaller $b$ or $d$ has larger
probability than the second power-law with the bigger $b$ or $d$.

\begin{figure} [h]
\includegraphics[scale=.45]{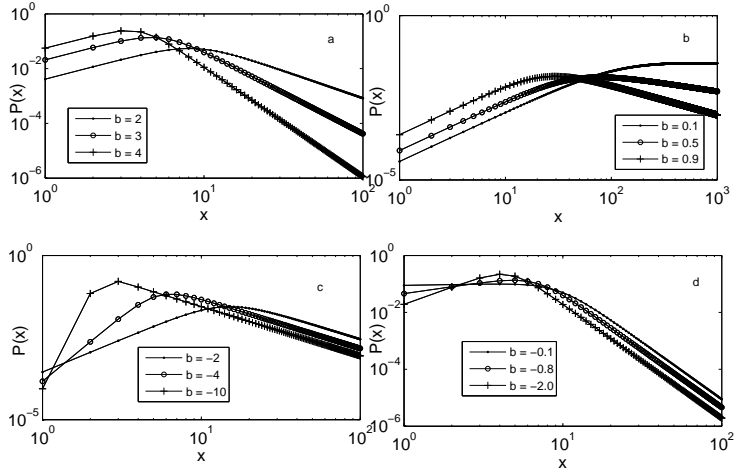}
\centering \caption{Variation tendency of probability function
$p(x)$ against parameter $b$ when one of $b, d$ is negative. From
each panel, we could find the exponents of the first power-law are
negative and those of the second power-law are positive. With the
increasing of the absolute value of parameter $b$, the exponent of
power-law depended on $b$ is increasing. The value of crossover
becomes smaller when the absolute value of parameter $b$
increases.} \label{pb1fig}
\end{figure}

Next, we will discuss the second class which means the function
exhibits nonmonotonicity like Figure \ref{pxfig}. In Figure
\ref{pb1fig}, we take the function as $p(x)=a/(x^b+1999*x^{-1.5})$
in the panels $a$ ($b>|d|$) and $b$ ($b<|d|$) with negative $d$.
In the panels $c$ ($|b|>d$) and $d$ ($|b|<d$) with negative $b$,
the function is $p(x)=a/(x^b+10^{-4}*x^{2.5})$. We could find the
variation tendencies are both the same for panel $a$ and $b$, and
for panel $c$ and $d$. For the former, as $b>d$, $b$ is in charge
of the second power-law. Meanwhile, because of $|b|>|d|$ in panel
a, probability function has small value of crossover. However, in
panel b, because of $|b|<|d|$, probability function has big value
of crossover. For the latter, as $b<d$, $d$ is in charge of the
second power-law. In panel c, since $|b|>|d|$, probability
function has small value of crossover. It is vice versa in panel
d. So, we draw a conclusion that the second power-law is charged
by bigger parameter out of $b$ and $d$. And, probability function
has big value of crossover when $|d|>|b|$.

\section{Clustering coefficient}
A common property of social networks is that cliques form,
representing circles of friends or acquaintances in which every
member knows every other member. This inherent tendency in cluster
is quantified by the clustering coefficient. For each node $i$
having $x_i$ edges which connect it to $x_i$ other nodes, if the
nearest neighbors of this node are part of a clique, there would
be $x_i (x_i-1)/2$ edges between them. The radio between the
number $E_i$ of edges that actually exist between these $x_i$
nodes and the total number $x_i (x_i-1)/2$ gives the value of the
clustering coefficient of node $i$:
\begin{equation}
C_i=\frac{2E_i}{x_i(x_i-1)}.
\end{equation}
The clustering coefficient of the whole network is the average of
all individual $C_i$'s. By calculation, $C$ of the entire urban
road network is 0.042. To check whether our network is a
small-world network or not, we construct a random network with
same number of nodes and edges in our network and get
$C_{rand}=0.004$ which is much smaller than that of urban road
network.

\section{Characteristic path length}
Shortest paths play an important role in the transport and communication
within a network. It is useful to represent all the shortest path lengths
of a graph $G$ as a matrix $D$ in which the entry $d_{ij}$ is the length
of the geodesic from node $i$ to node $j$. The maximum value of $d_{ij}$ is
called the diameter of the network. A measure of the typical separation between
two nodes in the network is given by the average shortest path length, also
known as characteristic path length, defined as the mean of geodesic lengths over
all couples of nodes \cite {SB}:
\begin{equation}
L=\frac{1}{N(N-1)}\sum_{i,j\in G,i \neq j}d_{ij}.
\end{equation}
From above equation, we get the average shortest path of urban
road network $L=4.183$ and $L_{rand}=4.000$ from artificial random
network we constructed. $L\approx L_{rand}$ and $C\gg C_{rand}$ suggest
that urban road network of Le Mans has the small-world network's properties.
In the meanwhile, we calculate the diameter of this network and its value is
equal to 8.

\section{Concluding remarks}
In the present paper, we have done a empirical study on urban road
network of Le Mans in France and investigated some structure
properties of it. From its degree distribution, double power-law,
we find that urban road network as one of spatial networks has
special degree distribution form from other networks, since there
are many constrains from its geographical embedding. The
phenomenon that the node with smallest degree has largest
probability in most of single power law distribution is not true
here. The node at the crossover point has the largest probability
in network we studied.

In order to make a deeply understanding of double power-law
distribution, we elaborately analyze the structure of the
probability function. Our discussions are divided into two classes
based on the feature of degree distribution of two transportation
networks, road network and airport network. One class is that both
of $b$ and $d$ are positive; another is that either of two
parameters is negative. Through our analysis, the conclusion is
that the bigger parameter of $b$ and $d$ is in charge of the
behavior in the second power-law with bigger degree. The smaller
one is in charge of that in the first power-law with smaller
degree. Absolute value of $b$ is more bigger than that of $d$, the
smaller the value of crossover. In addition, parameter $c$ just
controls the position of crossover in the limit of both $b$ and
$d$ unchanged.

At last, through the calculation of clustering coefficient and
average shortest path, small world properties have been found.
Though these topological properties are investigated and a novel
double power-law distribution function is given, the mechanism of
the organization of double power-law structure and that of the
transition from single power-law to double power-law are still
open questions in complex network study. It is necessary to
research more other features of double power-law distribution in
next stage, such as the behavior of its entropy. Perhaps a model
producing such distribution is worthwhile to consider.


\end{document}